\documentclass[letter]{aa}
\usepackage{graphicx}
\usepackage{natbib}
\usepackage{color}
\bibpunct{(}{)}{;}{a}{}{,} % AA STYLE
\begin{document}

\def\ms{M$_{\odot}$}
\def\zs{Z$_{\odot}$}
\def\mga{$^{24}$Mg}
\def\mgb{$^{25}$Mg}
\def\mgc{$^{26}$Mg}
\def\msun{${\rm M_{\odot}}$}
\def\cratio{$^{12}{\rm C}/^{13}{\rm C}$}

\title{Inhibition of thermohaline mixing by a magnetic field in Ap star descendants: Implications for the Galactic evolution of $^3$He}

\author{Corinne Charbonnel\inst{1,2} and Jean-Paul Zahn\inst{3}
        }

\authorrunning{C. Charbonnel and J.-P. Zahn}
 
\titlerunning{Magnetic inhibition of thermohaline mixing in red giant stars}

\offprints{C. Charbonnel}

\institute{ Geneva Observatory, University of Geneva, chemin des Maillettes 51, 1290 Sauverny, Switzerland \\
\email{Corinne.Charbonnel@obs.unige.ch}
\and
Laboratoire d'Astrophysique de Toulouse et Tarbes, CNRS UMR 5572, Universit\'e Paul Sabatier Toulouse 3, 14, av. E.Belin, 31400 Toulouse, France
\and
LUTH, Observatoire de Paris, CNRS, Universit\'e Paris-Diderot; Place Jules Janssen,     ~92195 Meudon, France\\
\email{Jean-Paul.Zahn@obspm.fr}    }
\date{Accepted for publication in A\&A Letter, October 30, 2007 - To appear in Vol.~476, Issue~3}

\abstract{}{To reconcile the measurements of $^3$He/H in Galactic HII regions with
high values of $^3$He in a couple of planetary nebulae, 
we propose that thermohaline mixing is inhibited by a fossil
magnetic field in red giant stars that are descendants of Ap stars. 
}
{We examine the effect of a magnetic field on the 
salt-finger instability, using a local analysis.}
{We obtain a threshold for the magnetic field of $10^4-10^5$ Gauss, above which
it inhibits thermohaline mixing in red giant stars located at or above the bump. 
Fields of that order are expected in the descendants of the Ap stars, 
taking into account the contraction of their core.}
{We conclude that in a large fraction of the descendants of Ap stars thermohaline mixing 
does not occur. As a consequence these objects must produce $^3$He as predicted 
by the standard theory of stellar evolution and as observed in the planetary 
nebulae NGC~3242 and J320. The relative number of such stars with respect to 
non-magnetic objects that undergo thermohaline mixing is consistent with the 
statistical constraint coming from observations of the carbon isotopic ratio 
in red giant stars. It also satisfies the Galactic requirements for the evolution
of the $^3$He abundance.}

\keywords{Instabilities; Stars: abundances, evolution; Galaxies: evolution}

\maketitle

\section{Introduction}

Numerous observations provide compelling evidence for a non-canonical 
mixing process which modifies the surface abundances of Li, C and N 
of low-mass red giants\footnote{i.e., stars with initial masses below $\sim$ 2 - 2.5~M$_{\odot}$
which evolve along the Red Giant Branch (RGB) to high luminosities until helium
is ignited in their core under degenerate conditions.} when they reach the bump 
in the luminosity function (see references in Charbonnel \& Zahn 2007, hereafter CZ07).
The quest for a mechanism which could lead to that extra-mixing 
between the hydrogen-burning shell and the convective envelope of 
red giant branch (RGB) stars has initially been unsuccessful. 
For example rotational mixing, i.e. rotation-induced meridional circulation 
and shear turbulence (Zahn 1992; Maeder \& Zahn 1998), is insufficient to explain 
the observed abundance pattern, as was shown by Charbonnel \& Palacios (2004) 
and Palacios et al.~(2006). But recently Eggleton et al.~(2006)
suggested a possible cause of such mixing, namely the molecular weight inversion 
created by the $^3$He($^3$He,2p)$^4$He reaction in the upper part of the hydrogen-burning shell. 
Based on 3D hydrodynamic simulations of a low-mass star at the RGB tip 
(Dearborn et al. 2006),  they found that such a $\mu$-profile is unstable and 
produces very efficient mixing. They claimed that this mixing was due to the 
well-known Rayleigh-Taylor instability, which occurs in incompressible fluids 
when there is a density inversion. In stellar interiors, which are stratified 
due to their compressibility, a similar dynamical instability occurs when 
the Ledoux criterion for convection is satisfied (Ledoux 1947),  
but it acts to render the temperature gradient adiabatic rather than to suppress 
the density inversion. Presumably it is this instability that 
Eggleton and colleagues observed with their 3D code, as attested 
by the high velocities they quote, because they started their simulation with 
a unrealistic 1D model which did not include any mixing, in which the $\mu$-inversion 
had already attained its maximum value.

In reality, the first instability to occur in a star, as the inverse $\mu$-gradient 
gradually builds up, is  the thermohaline instability, as was pointed out by 
CZ07. 
This double-diffusive instability is observed in salted water in the form of 
elongated fingers, when the temperature is stably stratified, but salt is not, 
with fresh water at the bottom and salted at the top, the overall stratification 
being dynamically stable (Stern 1960). 

It was Ulrich (1972) who first noticed that the $^3$He($^3$He,2p)$^4$He reaction 
would cause a $\mu$-inversion, and he was the first to derive a prescription 
for the turbulent diffusivity produced by the thermohaline instability in stellar 
radiation zones. This prescription is based on a linear analysis, and it is certainly 
very crude, but it has the merit to exist. When it is applied to the $\mu$-inversion 
layer in RGB stars, with the shape factor recommended by Ulrich and sustained 
by laboratory experiments (Krishnamurti 2003), it yields 
a surface composition that is compatible with the observations of the carbon isotopic 
ratio as well as of the abundances of lithium, carbon and nitrogen 
in RGB stars (Fig.~3 of CZ07).

Simultaneously thermohaline mixing leads to the destruction of most of the $^3$He 
produced during the star's lifetime (Fig.~4 of CZ07).
It seems thus that we have finally identified the mechanism responsible for 
abundance anomalies in RGB stars, which simultaneously accounts for the measurements
of $^3$He in Galactic HII regions (Balser et al.~1994, 1999a; 
Bania et al.~1997, 2002)\footnote{According to classical theory of 
stellar evolution which predicts a large production of $^3$He by low-mass stars, 
Galactic HII regions should be highly enriched in $^3$He since they formed 
out of matter that has undergone 12 billion years of chemical evolution. 
However, their $^3$He content is similar to that of the Sun, and thus no evidence 
for any enrichment of $^3$He during the last 4.5 Gyr was found. 
In addition, the ``best" determination in a Galactic HII region (namely S209) 
by Bania et al.~(2002) has yielded a $^3$He abundance almost identical to the WMAP value,
indicating that the stellar contribution to $^3$He is very small. }.

A problem remains however, namely that all RGB stars brighter than the bump should 
undergo thermohaline mixing since they develop the same $\mu$-inversion.
As underlined by Balser et al.~(2007), this was obviously not the case 
at least in two stars, i.e., in the RGB progenitors of the Planetary Nebulae NGC~3242 and J320. 
Slightly more massive than the Sun, these PNe are indeed caught in the act of returning to
the interstellar matter fresh elements synthesized in their womb among which
$^3$He with the amount predicted by classical stellar models (Balser et al.~1997, 
1999b, 2006; Galli et al.~1997).
On the other hand the carbon isotopic ratio data indicate that a few RGB stars escape that extra-mixing - 
about 4$\%$ according to Charbonnel \& do Nascimento (1998). 
This is consistent with the predictions of the chemical evolution models, which suggest that 
fewer than 10$\%$ of all planetary nebulae enrich the ISM with $^3$He at the level of 
classical stellar models (Galli et al.~1997; Tosi 1998, 2000;
Palla et al.~2000, 2002; Romano et al.~2003), thus supporting the self-consistency 
of the extra-mixing mechanism.

It is tempting to confront these numbers with the fraction of the Ap stars 
(suspected to all host magnetic fields) relative to all A stars, which 
is $5 - 10\%$ as estimated by Wolff (1968). Since A-type stars are the progenitors 
of the more massive RGB stars, we are led to conjecture that in those giants that 
show no sign of extra-mixing, the thermohaline instability is inhibited by a 
deeply buried magnetic field.

Here we check that possibility by examining the effect of a magnetic field on 
the salt-finger instability, using a simple local analysis.

\section{Thermohaline instability in the presence of a magnetic field}
\label{lin-anal}
All perturbations (displacement $\vec \xi$, magnetic field $\vec b$,
pressure $P'$, temperature $T'$, molecular weight $\mu'$) are expanded in Fourier modes
\begin{equation}
\exp[ i(k_x x+k_y y+k_z z) +s t],
\end{equation}
where we use local Cartesian coordinates, with the $z$-axis pointing in the vertical direction.

We begin by deriving the buoyancy force, which is weakened through radiative damping;
we neglect all other forms of dissipation that operate more slowly: Ohmic, molecular diffusion, viscosity.
We split the buoyancy frequency in two parts, the first due to the thermal stratification and the second to the composition gradient:
\begin{equation}
N^2 = N_t^2 + N_\mu^2  = {g \delta \over H_P} (\nabla_{\rm ad} - \nabla)
+  {g \varphi \over H_P} \nabla_\mu
\end{equation}
with the usual notations, and 
$$
\nabla_\mu  =  {d \ln \mu \over d \ln P} \qquad
 \delta = -\left(\partial \ln \rho \over \partial \ln T\right)_{\!P,\mu}  \quad  
\varphi=\!\left(\partial \ln \rho \over \partial \ln \mu \right)_{\!P,T} \!. \nonumber
$$
We shall consider here only the case where the stratification is dynamically stable ($N^2 \geq 0$), i.e. when the Ledoux criterion for stability is satisfied:
\begin{equation}
\nabla_{\rm ad} - \nabla + \left({\varphi \over \delta}\right)\nabla_\mu > 0 .
\label{ledoux}
\end{equation}

The linearized heat equation may be written as
\begin{equation}
s {T' \over T} + {N_t^2 \over g} \, s \xi_z = - \kappa k^2 {T' \over T} ,
\end{equation}
where  $k^2=k_x^2+k_y^2+k_z^2= k_h^2 + k_z^2$ and $\kappa$ is the thermal diffusivity.
(We simplify the Laplacian by assuming that $\kappa$ does not vary much over one vertical wavelength.)
Likewise, the advection equation for the molecular weight perturbation takes the form
\begin{equation}
s  {\mu' \over \mu} - {N_\mu^2 \over g} \, s \xi_z = 0 .
\end{equation}
Thus the buoyancy force is given by
\begin{equation}
- g {\rho' \over \rho} = g \left( {T' \over T} - {\mu' \over \mu}\right)=
- \left[{N_t^2 \over 1+ \kappa k^2/s} + {N_\mu^2 }\right]\xi_z .
\end{equation}

Turning next to the Lorentz force,
we perturb the magnetic field ${\vec B}_0=(B_x, 0, B_z)$ by the displacement $\vec \xi$, and draw the field perturbation $\vec b$ from the  induction equation:
\begin{equation}
{\vec b} = \vec \nabla \times ({\vec \xi} \times {\vec B}_0)  .
\end{equation}
It is then straightforward to calculate the perturbation of the Lorentz force per unit volume:
\begin{equation}
{\vec L} = {1 \over 4 \pi \rho} \left[ ({\vec \nabla} \times {\vec b}) \times {\vec B_0} 
\right] ,
\end{equation}
where we neglect the variation of ${\vec B}_0$ over a perturbation wavelength. From here on we drop the index $0$ designating the basic field.

It remains to implement the buoyancy and Lorentz forces in the equation of motion,  which we simplify by neglecting the effect of rotation.
\begin{equation}
s^2 {\vec \xi}+ \left[{N_t^2 \over 1+\kappa k^2/s} + N_\mu^2 \right] {\vec e}_z \xi_z = 
-i {\vec k} {P' \over \rho} +{\vec L}.
\end{equation}
Introducing the Alfv\'en velocity $\vec a = \vec B / \sqrt{4  \pi \rho}$, we get
\begin{eqnarray}
\label{eq-system}
&i& \!\!\! k_x  {P' \over \rho} + \left[s^2 +a_z^2 (k_x^2+k_z^2) \right] \xi_x 
 \\
&& +k_y a_z (k_x a_z -k_z a_x) \,\xi_y  - k_x^2 a_x a_z \, \xi_z
 = 0 \, , \nonumber  \\
&i& \!\!\!  k_y {P' \over \rho} + k_y (k_x a_z - k_z a_x) (a_z \xi_x - a_x \, \xi_z)  
\nonumber \\ 
&&  + \left[ s^2 +(k_x a_x + k_z a_z)^2 + k_y^2(a_x^2+a_z^2) \right] \xi_y = 0  \, ,\nonumber \\
&i& \!\!\!  k_z  {P' \over \rho} - a_x a_z (k_x^2+k_z^2) \, \xi_x - k_y a_x (k_x a_z - k_z a_x) \, \xi_y  \nonumber\\
&& +\left[s^2 + a_x^2(k_x^2+k_z^2) + {N_t^2 \over 1+ \kappa s^2 / \sigma} +  N_\mu^2\right]\xi_z  = 0 \, , \nonumber
\end{eqnarray}
which we complete with the continuity equation, written in its simplest form, assuming that the Boussinesq 
approximation is valid, i.e. that $|k_z| \gg |d \ln \rho / dz|$:
\begin{equation}
k_x  \xi_x + k_y \xi_y + k_z \xi_z =0 .
\label{cont-eq}
\end{equation}
The determinant of that fourth-order system is the dispersion relation from which we derive the eigenvalues $s$, i.e. the growth-rate of the instability.

In the non-dissipative case, this dispersion relation reduces to (cf. Kumar et al. 1999)
\begin{equation}
\Big( s^2 +  (\vec k \cdot \vec a)^2  \Big)
\Big( s^2 + {k_h^2 \over k^2} N^2 + (\vec k \cdot \vec a)^2 \Big)
 = 0 .
\end{equation}
The first set of (imaginary) roots corresponds to Alfv\'en waves, and the second 
to internal gravity waves modified by the magnetic field.

When thermal dissipation is included, these waves are damped, and a fifth root appears, 
which yields the growth-rate of the thermohaline instability in the presence of a magnetic field:
\begin{equation}
s = -\kappa k^2 \,  {N_\mu^2 \, (k_h/k)^2  + (\vec k \cdot \vec a)^2  \over
(N_t^2 + N_\mu^2)(k_h/k)^2  + (\vec k \cdot \vec a)^2 } .
\end{equation}
Note that the magnetic field always plays a stabilizing role; it overcomes the destabilizing effect of the inverse $\mu$ gradient when 
\begin{equation}
 (\vec k \cdot \vec a)^2  >  - N_\mu^2 \, (k_h/k)^2 .
\end{equation}

We may assume that the horizontal size ($\lambda=2 \pi / k_h$) of the $^3$He fingers is much smaller than their vertical extent: $k_h \gg k_z$. Then it is mainly the horizontal component $B_h$ of the field that prevents the thermohaline instability, when
\begin{equation}
B_h^2  > { \rho \, \lambda^2 \over \pi} | N_\mu^2 | .
\label{threshold}
\end{equation}
This criterion is sufficiently accurate for our purpose, although it has been established 
assuming that $| N_\mu^2 |$ does not vary much over a vertical wavelength of the perturbation, 
a condition which is not fulfilled in the real star.

\begin{figure}
\centering
\resizebox{0.7\hsize}{!}{\includegraphics{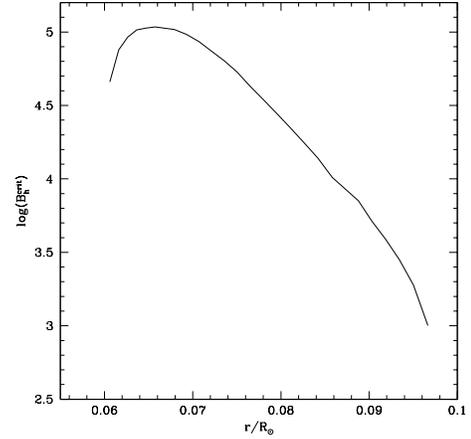}}
\caption{Critical value of the magnetic field which inhibits thermohaline mixing 
in a 1.5 M$_{\odot}$, Z$_{\odot}$ model located at the bump.}
\label{profilsabond}
\end{figure}

We are aware that the magnetic field itself may undergo instabilities 
(see Spruit 1999; Menou et al.~2004) that are not captured by our crude analysis, 
since we ignored both rotation and the spatial variation 
of ${\vec B}_0$\footnote{Busso et al.~(2007) suggested that such instabilities might be 
the cause of extra-mixing. In that case however, the presence of deeply buried magnetic 
fields would have to be the rule, rather than the exception. We feel thus that the scarcity of magnetism 
observed in the A-type stars speaks in favor of the alternate scenario that we present here.}.
We have thus implicitly assumed that magnetic instabilities, if any, 
appeared before the inverse $\mu$-gradient built up, and that they led to a stable field configuration.

Figure~1 shows the critical value of $B_h$ predicted by Eq.~(\ref{threshold}) 
in the region of $\mu$-inversion of a 1.5 M$_{\odot}$ model with solar metallicity.
This model is located at the bump in the luminosity function,
i.e., where thermohaline mixing should be fully efficient in the absence of magnetic field
and where the C, N and Li abundance anomalies show up on the RGB (Charbonnel 1994).
Here we assume that the finger width is $1/10$ of the height of the
$\mu$-inversion region.
According to our criterion, a magnetic field of $B_h~\approx~10^5$ Gauss suppresses entirely
the thermohaline mixing, while a field of $10^4$G is sufficient to inhibit 
the instability in the upper 1/3 of the $\mu$ inversion domain.
We also computed (not shown here) the critical value of $B_h$ at the bump
in a 2 M$_{\odot}$ model at solar metallicity, and also in a 0.9 M$_{\odot}$ model
with [Fe/H]=-1.8. Due to the fact that the structure of the H-burning region is very similar
in all the cases, we obtain comparable values for $B_h^{crit}$.

Magnetic fields of that order are expected in the deep interior 
of the red giant descendants of the Ap stars.
On the main sequence, these Ap stars possess surface fields of a few $10^2$ to 
about $3\times 10^4$G, which are believed to be of fossil origin; 
if so, the field is deeply rooted and its Ohmic decay time largely exceeds the main sequence lifetime. 
This scenario seems to be confirmed by a recent investigation which concludes 
that Ap stars whose mass is less than 3~M$_{\odot}$ show no compelling evidence
for field decrease during their main sequence life 
(Landstreet et al.~2007)\footnote{The rapid field decline observed above this mass may be 
explained by the increased turbulent diffusion occurring in the larger convective core.}. 
The hypothesis of the fossil nature of magnetism in the Ap stars has recently 
been strengthened by the discovery that their progenitors are certainly 
the Herbig Ae/Be stars (Wade et al.~2007a, b).

After the main sequence phase, as the central region of the star contracts 
during evolution on the giant branch, the magnetic field may be enhanced 
by up to two orders of magnitude, due to flux conservation. Thus it can easily 
exceed the critical strength that ensures the suppression of the thermohaline instability.
We thus consider as plausible that those red giants which are 
prevented from undergoing the non-canonical mixing are the descendants 
of main sequence Ap stars, because the magnetic field of fossil origin suppresses 
in them the thermohaline instability.
If this scenario is correct, one should observe magnetic fields in 
RGB stars that behave classically, and in PNe such as NGC~3242 and J320.

\section{Discussion}

The observational uncertainties in effective temperature, luminosity and 
metal content of Ap stars hamper a precise determination of their mass (see Bagnulo et al.~2006).
According to a survey covering the Ap/Bp stars located within 100 pc from the Sun 
(Power et al.~2006), magnetic stars range at solar metallicity 
between 3.6 and 1.5~M$_{\odot}$  (with one exception at 6~M$_{\odot}$). 
The lower limit coincides with the appearance of developed surface convection zones,
which depends on the initial stellar metallicity.
If this correlation held in the past independently of the bulk stellar metallicity, metal-poor 
magnetic stars could have extended to lower initial mass: standard models predict indeed 
that the convection zone of a 0.9 M$_{\odot}$ main sequence star with [Fe/H]=-2 
has the same thickness as that of a 1.5 M$_{\odot}$ main sequence star of solar metallicity 
(Talon \& Charbonnel 2004).
As a consequence, one can conjecture that below solar metallicity, Ap star counterparts  
formed with  initial masses much below 1.5~M$_{\odot}$.
This expected dependancy of the mass threshold for the magnetic inhibition of 
thermohaline mixing with metallicity is of direct relevance to the Galactic evolution 
of $^3$He and will have to be taken into account in future chemical evolution models 
which consider metal-dependant yields.

Galli et al.~(1997) have determined a mass of 1.2~$\pm 0.2$~M$_{\odot}$  
for the progenitor of the planetary nebulae NGC~3242. 
Although such a determination is quite a feat and is undermined by several 
observational and theoretical difficulties, we find very encouraging 
that the initial mass derived for this object is so close to the lower mass limit 
for Ap stars at solar metallicity. We have to note indeed that the metallicity 
of NGC~3242 is half-solar (Barker 1985), so that the lower mass for Ap stars 
may already be shifted towards a lower value.

It remains to be checked whether the magnetic stars are in sufficient number 
on the main sequence to account for the $\approx 4\%$ of red giants which are spared 
from non-canonical mixing as estimated from the carbon isotopic ratio data 
(Charbonnel \& Do Nascimento 1998). 
The classical value of the incidence of Ap/Bp stars, which is $5-10\%$ 
based on a magnitude-limited sample (Wolff 1968), has been challenged 
by the aforementioned survey performed by Power et al.~(2006): 
in the solar neighborhood they find only 1.7\% of magnetic stars. 
This value is close to that (i.e., $\sim 3.5 \%$) derived by North (1993) for 
Ap stars with masses between 1.7 and 2.5~M$_{\odot}$ that are members of Galactic open clusters,
and by Paunzen et al.~(2006; $\sim 2.2 \%$) for chemically peculiar stars in 
young LMC open clusters. 
We note that the bulk incidence of Ap stars is 
very sensitive to the chosen value of the lower mass limit for these objects 
in a given sample (see Fig.~4 of Power et al.~2006 or Table VII of North~1993).
In conclusion, we consider that the present statistics 
support our suggestion that thermohaline mixing is inhibited by a fossil magnetic field
in those red giant stars that are the descendants of main sequence Ap stars.
Within this framework we reconcile the long-standing problem of $^3$He overproduction 
on the Galactic timescale with the high values of $^3$He observed in a couple of 
planetary nebulae. 
We encourage the observers to look for magnetic fields in ``classical" RGB 
and in NGC~3242 and J320 which we propose to call ``thermohaline 
deviant stars".

\begin{acknowledgements}
We thank the referee for her/his helpful comments.
CC is particularly grateful to D. Balser, T. Bania, C. Chiappini, 
R. Rood and M. Tosi for enlightening discussions on the ``$^3$He problem"  
over the years. We thank J. Landstreet, P. North and G. Wade for their
comments on Ap statistics. 
CC is supported by the Swiss National Science Foundation (FNS). 
We acknowledge the financial support of Programme National de
Physique Stellaire (PNPS) of CNRS/INSU, France.
\end{acknowledgements}

{}

\end{document}